\pdfoutput=1
\documentclass[aps,prl,amsmath,amssymb,twocolumn,superscriptaddress,notitlepage,reprint]{revtex4-1}
\usepackage{bbm}
\usepackage{amsmath}
\usepackage{verbatim}
\usepackage{graphicx}
\usepackage{times}
\usepackage{amssymb}
\usepackage{epsfig}
\usepackage{graphicx}
\usepackage{bm}
\usepackage{times}
\usepackage{txfonts}
\usepackage{color}
\usepackage[normalem]{ulem}

\newcommand{\CSL}{\mathrm{CSL}}
\newcommand{\DP}{\mathrm{DP}}
\newcommand{\diff}{\mathrm{d}}
\newcommand{\vk}{\boldsymbol{k}}
\renewcommand{\vr}{\boldsymbol{r}}
\newcommand{\vq}{\boldsymbol{q}}
\newcommand{\vG}{\boldsymbol{G}}
\newcommand{\cL}{ {\cal L} }
\newcommand{\Op}[1]{\mathsf #1}

\newcommand{\ox}{ \Op{x} }
\newcommand{\op}{ \Op{p} }
\newcommand{\sinc}{{\rm sinc}}
\newcommand{\erf}{{\rm erf}}
\newcommand{\tr}{{\rm tr}}
\newcommand{\vs}{\boldsymbol{s}}
\newcommand{\ovr}{ \boldsymbol{\Op{r}} }
\newcommand{\ovR}{ \boldsymbol{\Op{r}} }
\newcommand{\om}{ \Op{m} }

\begin{document}

\title{Optomechanical sensing of spontaneous wave-function collapse}

\author{Stefan Nimmrichter}

\affiliation{Faculty of Physics, University of Duisburg-Essen, Lotharstra\ss{}e 1, 47048 Duisburg, Germany}

\author{Klaus Hornberger}

\affiliation{Faculty of Physics, University of Duisburg-Essen, Lotharstra\ss{}e 1, 47048 Duisburg, Germany}

\author{Klemens Hammerer}

\affiliation{Institute for Theoretical Physics and Institute for Gravitational Physics (Albert Einstein Institute), Leibniz University Hannover, Callinstra\ss{}e 38, 30167 Hannover, Germany}

\date\today

\begin{abstract}
 Quantum experiments with nanomechanical oscillators are regarded as a testbed for hypothetical modifications of the Schr\"{o}dinger equation, which predict a breakdown of the superposition principle and induce classical behavior at the macro-scale.  It is generally believed that the sensitivity to these unconventional effects grows with the mass of the mechanical quantum system.
 Here we show that the opposite is the case for optomechanical systems in the presence of generic noise sources, such as thermal and measurement noise. We determine conditions for distinguishing these decoherence processes from possible collapse-induced decoherence in continuous optomechanical force measurements.
\end{abstract}

\maketitle

\paragraph*{Introduction.}The observation of quantum behaviour in a growing number of macroscopic systems of light or matter has demonstrated the validity of the 
superposition principle at impressively large scales \cite{Hammerer2010,RevModPhys.84.1765,RevModPhys.84.777,arndt_testing_2014}. Within the framework of quantum mechanics the disappearance of coherent superposition states at macroscopic scales is attributed entirely to the interaction with uncontrolled and unobserved degrees of freedom. This is the paradigm of 
decoherence theory \cite{joos_decoherence_2003,Zurek2003,Schlosshauer2005}. According to alternative approaches, quantum theory must be modified at a fundamental level
to explain the emergence of 
macroscopic realism \cite{Leggett2002} 
and to solve the measurement problem \cite{Bassi2003}. 
Such modifications are designed 
to induce an objective collapse of the wave function above a critical 
mass scale of a given quantum system, thereby restoring classicality.

The most widely studied modification is the model of continuous spontaneous localization (CSL) \cite{Ghirardi1990b,Bassi2003,Bassi2013}, which introduces 
a nonlinear stochastic addition to the Schr\"odinger equation. The delocalized wave function of a massive particle gets gradually and randomly localized down to a microscopic length scale of about $100\,$nm, at a rate that amplifies with the particle's mass. In many respects, the CSL model can be regarded as the prime example of a broad generic class of macrorealist modifications; it is compatible with all experimental observations to date and with most of the symmetry 
principles underlying both quantum and classical mechanics \cite{Bassi2003,Adler2004,Nimmrichter2013}. Another important macrorealist model is the Di\'osi-Penrose  (DP) collapse mechanism \cite{Diosi1987,Diosi1989,Penrose1996,Bassi2013} which explains the effect by gravitational self-interaction.

The main prediction of these models is the objective breakdown of the quantum superposition principle with growing mass. This would be directly observable by a specific mass-dependent loss of visibility in interference experiments with 
nanoparticles \cite{Marshall2003,Romero-Isart2011b,Nimmrichter2011a,Hornberger2012_RMP,Bateman2013}. However, the 
mass range where macrorealistic collapse should become effective has yet to be reached in experiments.

At the same time, the random localization process predicted by the CSL (and the DP) model inherently implies momentum diffusion, i.e.~a Brownian-like background noise, which also affects the classical motion of macroscopic matter. No quantum coherence is required to detect these hypothetical sources of noise \cite{Bahrami2014}. In fact, the necessary tools are being used 
in optomechanics labs worldwide: nano- and micromechanical oscillators manipulated and read out by optical fields are the most sensitive measurement devices for forces \cite{Clerk2010a,Aspelmeyer2013,Danilishin2012,Chen2013}. Recent experiments have demonstrated measurement sensitivities limited by radiation-pressure back-action noise~\cite{Murch2008,Safavi-Naeini2012,Purdy2013,Matsumoto2013}, a necessary condition for reaching the standard quantum limit of continuous force measurements~\cite{Caves1980a,Caves1981}.
Such measurements 
can give rise to Gaussian entangled states of macroscopic masses
\cite{PhysRevLett.100.013601} which can be used to test DP models \cite{PhysRevA.81.012114}.

In this letter, we assess the requirements for detecting the force noise postulated by macrorealistic models in optomechanical devices.
We show that ultra-sensitive force measurements at cryogenic temperatures using oscillators with low (sub-Hz) resonance frequencies, sub-gram masses, and high mechanical quality factors can test significant and unexplored parameter regimes of those models. Surprisingly, we find that \emph{higher} masses will in general imply \emph{lower} sensitivities.

Our results serve also as a benchmark for optomechanical superposition experiments \cite{Marshall2003,Romero-Isart2011b,Nimmrichter2011a,Bateman2013}. The tiny oscillation amplitudes of such oscillators (on the scale of femto- to picometers) allow the collapse modification to be approximated by a linear diffusion term. The coherence loss rate for all superposition states is therefore determined by the same momentum diffusion rate that governs motional noise \cite{Bassi2005}. In the following, we  focus on the CSL model, and state the analogous DP results only in the end.

\paragraph*{Noise induced by localization.} The CSL model is characterized by two parameters, the localization length $r_\CSL$ and the localization rate $\lambda_\CSL$. The former, conventionally set to the value $r_\CSL=100\,$nm, determines the size down to which delocalized quantum states get localized. The latter gives the average localization rate at one proton mass; it is currently believed not to exceed $10^{-8 \pm 2}\,$Hz \cite{Fu1997,Adler2007a,Bassi2013}. 
%SN: Replaced 'in the range' by 'not to exceed' 
%SN: Added citations to Adler2007a and Fu1997.
The challenge is to identify experimental testbeds where the localization rate $\lambda_\CSL$ can be sensed at this level \cite{Feldmann2012}.
%SN: Added cite to Feldmann2012 (estimated bounds on CSL by experiments)

The effective localization and diffusion rates of many-body systems amplify with mass, and they depend on the spatial extension of both the quantum state of motion and the mass distribution. Two regimes can be considered: For nanoparticles smaller than $r_\CSL$ prepared in superposition states with separations larger than $r_\CSL$, as relevant to matter-wave interferometry, the effective coherence loss rate grows in proportion to the squared mass \cite{Bassi2003,Adler2007,Nimmrichter2011a}. On the other hand, for micrometer-sized mechanical resonators delocalized over amplitudes much smaller than $r_\CSL$, as relevant 
in optomechanics, 
we find a \emph{sublinear} mass scaling of the effective localization and diffusion rate. To be specific, let us focus on 
cantilever configurations, where the center of mass of a rigid body of mass $m$ oscillates linearly, say, along the $x$-axis with an amplitude $x_0 \ll r_\CSL$. This could be a cubic mirror \cite{Marshall2003}, an optically 
trapped nanosphere \cite{Chang2010,Romero-Isart2010} 
or a micromembrane \cite{Thompson2008,Teufel2011}. In this limit, the observable consequences of the CSL model are approximated by a quantum master equation $\dot\rho=(\cL+\cL_\CSL) \rho$ where $\cL$ is a Liouvillian associated to standard quantum mechanics. The Lindblad term  $\cL_\CSL\rho = -D_\CSL \left[ \ox,\left[ \ox, \rho \right] \right]/\hbar^2$, with $\ox$ the center-of-mass position operator, describes the momentum diffusion implied by the CSL modification. It can be viewed as arising from a stochastic force $f_\CSL (t)$ characterized
by the two-time correlation function $\langle f_\CSL(t)f_\CSL(t')\rangle = D_\CSL\delta(t-t')$.
The associated diffusion rate $D_\CSL = \lambda_\CSL \left(\hbar/r_\CSL\right)^2 \alpha$ involves a mass-dependent geometry factor 
%SN: cited supplement once again.
\cite{Supplement},
\begin{align}
  \alpha &= \frac{r^5_\CSL}{\pi^{3/2} \mathrm{amu}^2} \int \diff^3 k \, k_x^2 e^{-r^2_\CSL \vk^2} \left| \tilde{\varrho} (\vk) \right|^2.  \label{eq:alpha}
\end{align}
Here, $\tilde{\varrho} (\vk) = \int\diff^3 r \, \varrho(\vr) e^{-i\vk\cdot\vr}$ denotes the Fourier transform of the mass density, normalized to the total mass, $\tilde{\varrho}(0)=m$. The cantilever can safely be described as a homogeneous rigid body since the dynamics and structure of the underlying crystal lattice vary well below the scale of $r_\CSL = 100\,$nm. Simple expressions for the geometry factor (\ref{eq:alpha}) are then obtained for materials with a constant mass density $\varrho$ and dimensions greater than $r_\CSL$ \cite{Supplement}. 
In the case of spheres and cubes with radii $R\gg r_\CSL$ and side lengths $b \gg r_\CSL$, we find
\begin{equation}
  \alpha_{\rm sph} (R) \approx \frac{16\pi^2 \varrho^2 r^4_\CSL}{3 \, \mathrm{amu}^2} R^2, \quad \alpha_{\rm cube} (b) \approx \frac{8\pi \varrho^2 r^4_\CSL}{\mathrm{amu}^2} b^2, \label{eq:alphasphcube}
\end{equation}
both proportional to $m^{2/3}$ at fixed $\varrho$. For the motion of a thin membrane 
(thickness $d\ll r_\CSL$, radius $R\gg r_\CSL$) along its symmetry axis, we obtain
\begin{align}
  \alpha_{\rm disc} (R) &\approx \frac{ 2 \pi^2 \varrho^2 r_\CSL^2 }{\mathrm{amu}^2} d^2 R^2. \label{eq:alphadisc}
\end{align}
Keeping the density and thickness fixed, the radius scales like $R = \sqrt{m/\pi \varrho d}$, and the geometry factor is at best proportional to mass. 
Exact geometry factors are derived in the Supplements \cite{Supplement,Note1}.
\begin{figure}[t]
  \centering
  \includegraphics[width=\columnwidth]{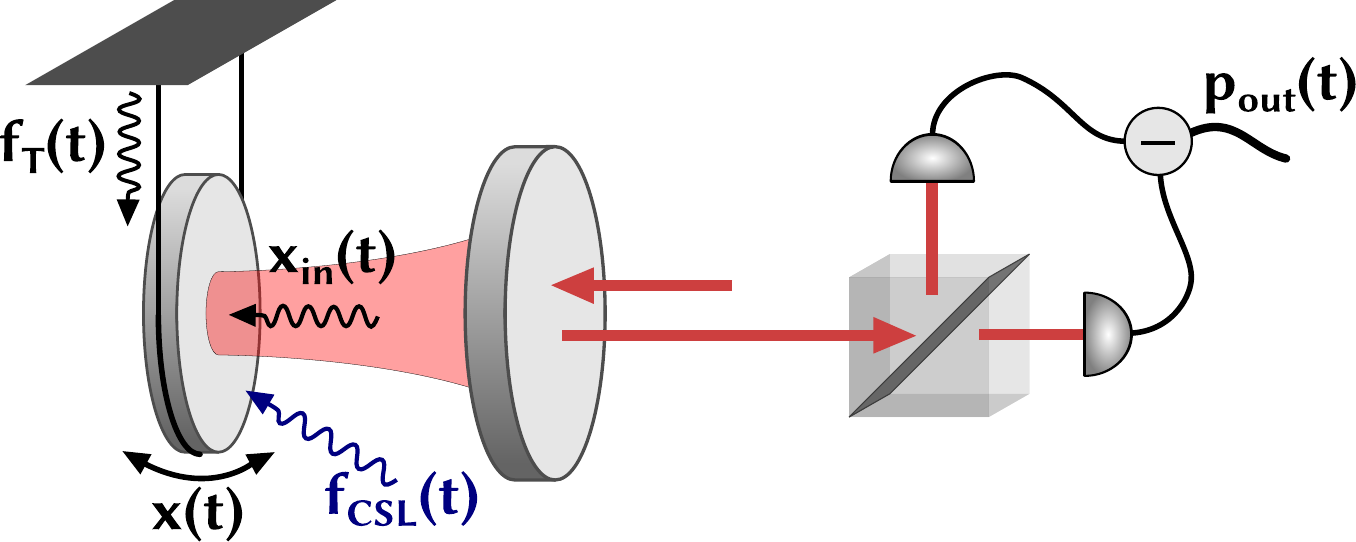}
  \caption{Sketch of an optomechanical setup for measuring macrorealistic noise forces. The harmonic motion $x (t)$ of a cantilever, monitored by means of an optical cavity field, is subject to thermal noise $f_T (t)$, optical amplitude noise $x_{\rm in} (t)$, and a hypothetical collapse-induced noise force $f_\CSL (t)$. All three contributions are reflected in the noise spectrum of the phase quadrature $p_{\rm out} (t)$ of the outgoing light field, as monitored by homodyne detection.}\label{fig:sketch}
\end{figure}

\paragraph*{CSL in the presence of thermal diffusion.} The overall sublinear  increase of the geometry factors with mass must be put into perspective by comparing CSL diffusion to standard sources of noise, most prominently, thermal noise. The mechanical motion in a thermal environment, and including the CSL effect, is sketched in Fig.~\ref{fig:sketch}. It is described by the Langevin equations of motion
\begin{align}
\dot{x}&=p/m, &
\dot{p}&=-m\Omega^2 x-\gamma p+f_T(t)+f_\CSL(t), \label{eq:mec_th}
\end{align}
where $\Omega$ and $\gamma$ are the mechanical resonance frequency and line width, respectively. The thermal noise force $f_T(t)$ is characterized by $\langle f_T(t)f_T(t')\rangle = D_T\delta(t-t')$ and $D_T = 2\gamma m k_BT$, 
valid in the relevant high-temperature limit $k_{\rm B}T\gg \hbar\Omega$ of the environmental bath.

The CSL momentum diffusion would dominate over thermal diffusion, and would thus be detectable in the noise spectrum,  
if $D_\CSL > D_T$, i.e.,
\begin{align}
\Lambda_T\equiv
\frac{2 r^2_\CSL \gamma k_B T}{\hbar^2} \, \frac{m}{\alpha} < \lambda_\CSL. \label{eq:thBound}
\end{align}
This condition gives a lower bound $\Lambda_T$ of CSL rate parameters for which the localization effect would be observable on top of the thermal noise spectrum; a significant test should aim for values between $0.1\,$nHz and 1\,$\mu$Hz.

Quite remarkably, the generally sublinear mass scaling of the geometry factors implies that more sensitive tests of the rate parameter require \emph{smaller} test masses (as long as the object stays larger than $r_\CSL=100\,$nm). This is in strong contrast to the quadratic enhancement of CSL detectability in conventional matter-wave interferometry \cite{Bassi2003,Nimmrichter2011a}, where 
the particle is smaller than the delocalization of its motional quantum state.
Moreover, our results illustrate that, for truly macroscopic bodies, the CSL scaling is carefully balanced: it rapidly restores classicality, while 
going practically unnoticed in the presence of a thermal environment.

\begin{figure}[t]
  \centering
  \includegraphics[width=\columnwidth]{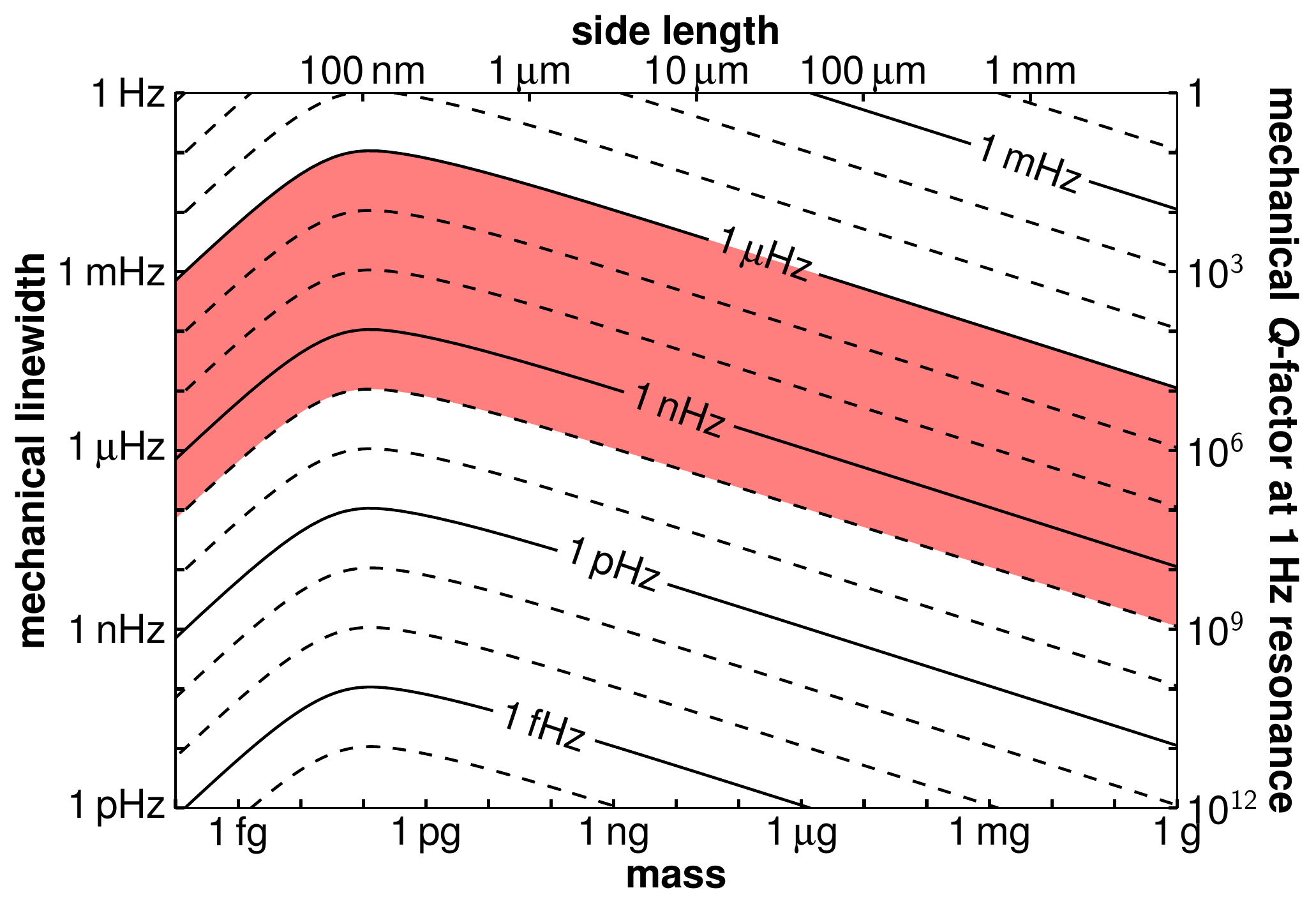}
  \caption{Lower bound $\Lambda_T$ for the detectable CSL rate parameter $\lambda_\CSL$ due to thermal noise at 1\,K, for varying mass $m$ and linewidth $\gamma$ of a cubic silicon cantilever, see Eq.~(\ref{eq:thBound}). The side length of the cube is given on the top, the quality factor $Q$ of a 1\,Hz oscillator on the right. The currently estimated upper bound for $\lambda_\CSL$ ($0.1\,$nHz to 1\,$\mu$Hz) is indicated by the shaded area.}\label{fig:lambda}
  %SN: replaced currently estimated 'range' by 'upper bound'.
\end{figure}

It is clear from (\ref{eq:thBound}) that a narrow linewidth (i.e.~a low frequency $\Omega$ and a high quality factor $Q = \Omega / \gamma$) is crucial for observing CSL noise against the thermal background. In addition, one must maintain a low temperature of the environment and monitor it precisely and independently of the noise level.
Figure\,\ref{fig:lambda} illustrates which masses $m$ and linewidths $\gamma$ are required to test given values of the localization rate $\lambda_\CSL$. We choose a  cubic silicon oscillator (side length $b$, mass density  $\varrho=2300\,$kg/m$^3$)
at the temperature $T=1\,$K.
With a $\Omega=1\,$Hz, $Q=10^6$, $1\,\mu$g-oscillator one could test $\lambda_\CSL$-values as small as $1\,$nHz, which matches 
currently estimated bounds. 
%SN: replaced 
%the current CSL estimates. 
The best sensitivity is obtained with oscillators of roughly the same size as $r_\CSL$. Since the approximation (\ref{eq:alphasphcube}) for the the geometry factor $\alpha_{\rm cube}$ fails in this case, we used the exact expression given in \cite{Supplement}.

\paragraph*{Effect of measurement noise.} Thermal noise is not the only limitation for detecting collapse-induced diffusion. The measurement process itself contributes back-action and shot noise to the readout signal~\cite{Clerk2010a,Aspelmeyer2013,Danilishin2012,Chen2013}. In optomechanics the mechanical resonator acts as a refractive (or reflective) element for optical fields. Its quadratures can thus be monitored by coupling the mechanical mode to a driven high-finesse cavity light mode.
The momentum diffusion of the oscillator can then be inferred from a continuous interferometric measurement of its position~\cite{Clerk2010a}. In the simplest scenario, as sketched in Fig.~\ref{fig:sketch}, the phase quadrature of a light field (relative to its coherent steady-state amplitude) will receive a signal linear in the oscillator's position, $p_\mathrm{out}(t)=p_\mathrm{in}(t)+g x(t)$.
Here $p_\mathrm{in}(t)$ is white measurement shot noise, $\langle p_\mathrm{in}(t)p_\mathrm{in}(t')\rangle=\frac{1}{2}\delta(t-t')$.
The strength of position transduction (of dimension Hz$^{1/2}$m$^{-1}$) is given by $g=k\sqrt{\mathcal{F}\Phi}$ in a typical optomechanical setup, where $k$ is the wave number of light, $\mathcal{F}$ is the finesse of the cavity, and $\Phi=P/\hbar\omega_\mathrm{opt}$ is the photon flux for a power $P$ injected into the interferometer~\cite{Clerk2010a}.

The mechanical oscillator will in turn be affected by a measurement back-action force proportional to the amplitude quadrature fluctuations $x_\mathrm{in}(t)$ of the light (again white noise), so that the 
momentum Langevin equation (\ref{eq:mec_th}) becomes
\begin{equation}\label{eq:mec}
\dot{p} = -m\Omega^2 x-\gamma p+f_T(t)+f_\CSL(t)+\hbar g x_\mathrm{in}(t).
\end{equation}
The Fourier components of the phase quadrature measured in the outgoing light field are then
\begin{align}
  p_\mathrm{out}(\omega)&=p_\mathrm{in}(\omega)+g\chi(\omega)\left[f_T(\omega)+f_\lambda(\omega)+\hbar g x_\mathrm{in}(\omega)\right],
\end{align}
where the mechanical susceptibility at measurement frequency $\omega$ takes the known Lorentzian form, $m \chi(\omega) = (\Omega^2-\omega^2+i\omega\gamma)^{-1}$.
From the measured phase quadrature we can infer the
total diffusion force at its Fourier frequency $\omega$,
\begin{equation}
f(\omega)=\frac{p_\mathrm{out}(\omega)}{g\chi(\omega)} = f_\CSL(\omega)+f_T(\omega)+\frac{p_\mathrm{in}(\omega)}{g\chi(\omega)}+\hbar gx_\mathrm{in}(\omega) .
\end{equation}
In this expression, the CSL-induced diffusion force competes with three unavoidable contributions to the signal fluctuation: thermal noise, shot noise, and measurement back-action noise.
The corresponding noise spectral density, 
as obtained from the recorded 
spectrum of the measurement signal, is
\begin{align}
S_{f}(\omega)& = D_{\CSL} + 2\gamma mk_BT+\frac{1}{2g^2|\chi(\omega)|^2}+\frac{\hbar^2g^2}{2}\nonumber\\
&\geq D_{\CSL} + 2\gamma mk_BT+\hbar |\chi(\omega)|^{-1}.\label{eq:SQL}
\end{align}
Here, the bound is achieved through optimization with respect to $g$ (i.e.~laser power $P$); it represents the standard quantum limit (SQL) of the continuous force measurement~\cite{Clerk2010a,Aspelmeyer2013,Danilishin2012,Chen2013}, achieved for $g_\mathrm{SQL}=1/\sqrt{\hbar |\chi(\omega)|}$.

As discussed above, it will be advantageous to work with a low frequency (Hz or sub-Hz), high-$Q$ oscillator to minimize the thermal noise contribution. For suitable measurement frequencies in the kHz range, we can therefore assume $\omega\gg\Omega,\gamma$, so that the high-frequency (free-mass) limit of the susceptibility applies, $|\chi(\omega)|\simeq 1/m\omega^2$. Note that this corresponds to the standard limit considered in gravitational wave detection~\cite{Danilishin2012,Chen2013}. In order to have the CSL diffusion dominate over the measurement-induced diffusion, the first term in \eqref{eq:SQL} must be larger than the last SQL term,
\begin{align}
\Lambda_{\rm SQL}\equiv\frac{ r^2_\CSL \omega^2}{\hbar} \, \frac{m}{\alpha} < \lambda_\CSL, \label{eq:mBound}
\end{align}
for the considered measurement frequency $\omega$. As in the case of the thermal bound (\ref{eq:thBound}), this lower bound for detectable CSL rate parameters increases with mass due to the sublinear mass scaling of the geometry parameter $\alpha$.

\begin{figure}[t]
  \centering
  \includegraphics[width=\columnwidth]{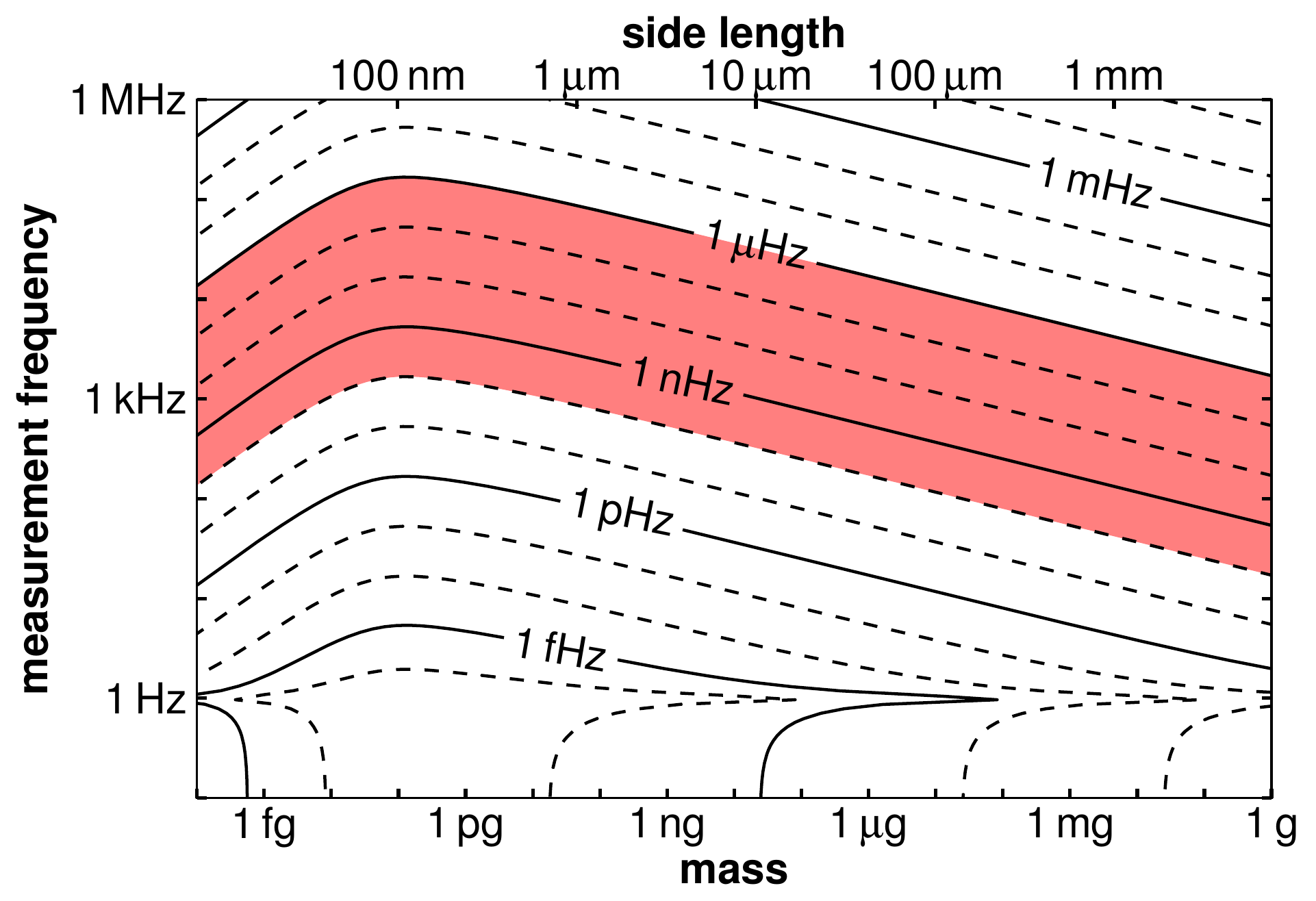}
  \caption{Lower bound $\Lambda_{\rm SQL}$ for the detectable CSL rates $\lambda_\CSL$ due to measurement noise 
  at SQL, as a function of mass $m$ and measurement frequency $\omega$, see Eq.~(\ref{eq:mBound}). We assume a cubic silicon cantilever (side length on the top) with frequency $\Omega=1\,$Hz and quality factor $Q=10^6$. Relevant upper bounds for the CSL rate are indicated by the shaded area.}\label{fig:lambda_SQL}
  %SN: replaced relevant 'values' by 'upper bounds'.
\end{figure}

Figure~\ref{fig:lambda_SQL} illustrates
when  CSL diffusion will be detectable on top of measurement noise, at  given measurement frequencies $\omega$ and masses $m$ of a cubic cantilever.
For a $\mu$g-oscillator, sensitivities down to $\lambda_\CSL>\Lambda_{\rm SQL}\simeq 1\,$nHz can be maintained at measurement frequencies up to hundreds of Hz.
The SQL coupling frequency in the free-mass limit, $g_\mathrm{SQL}=\omega\sqrt{m/\hbar}$, is equivalent to a laser power $P_\mathrm{SQL}=mc^2\omega^2/\mathcal{F}\omega_\mathrm{opt}$. 
Several recent experiments demonstrated back-action noise-limited detection with optomechanical systems~\cite{Murch2008,Safavi-Naeini2012,Purdy2013}, including a mg-scale mirror \cite{Matsumoto2013}. The SQL 
will eventually be reached by further reducing thermal background noise,
and possibly be overcome using Heisenberg-limited measurement strategies \cite{Danilishin2012}.

In summary, both the thermal bound (\ref{eq:thBound}) and the measurement bound (\ref{eq:mBound}) must be taken into account when probing CSL diffusion in optomechanical systems. The sum of both sets the achievable sensitivity in a given setup at SQL. In Table \ref{tab:CSL} we list the sensitivities reachable 
in a number of existing, proposed, and hypothetical configurations. We find that the current experimental state of the art is yet incapable of testing the CSL model at the relevant parameter range of $\lambda_\CSL < 10^{-8 \pm 2}\,$Hz. 
%SN: replaced '\sim' by '<'
To reach the desired sensitivity, one must aim for substantially improved quality factors and temperature control, rather than for high masses.

\begin{table*}
\begin{center}
\begin{tabular}[c]{lcccccccc}
\textbf{System} & $\varrho$ (g/cm$^3$) & $m$ & $\Omega/2\pi$ (Hz) & $Q$ & $T$ (K) & $\omega/2\pi$ (Hz) & $\Lambda_{T}$ (Hz) & $\Lambda_{\rm SQL}$ (Hz) \\ \hline
gravitational wave detector \cite{PhysRevLett.100.013601} & $2.3$ & $40\,$kg & $1$ & $25000$ & $300$ & $1000$ & $2\times 10^{-1}$ & $3\times 10^{-4}$\\
suspended disc \cite{Matsumoto2013} & $2.0$ & $5\,$mg & $0.5$ & $5 \times 10^5$ & $300$ & $500$ & $5\times 10^{-6}$ & $1\times 10^{-7}$\\
hypothetical setup 
& $2.0$ & $100\,\mu$g & $0.1$ & $10^6$ & $0.2$ & $100$ & $2\times 10^{-10}$ & $2\times 10^{-9}$\\
SiN membrane \cite{Purdy2013} & $3.4$ & $34\,$ng & $1.6 \times 10^6$ & $1100$ & $4.9$ & $1.6 \times 10^6$ & $4\times 10^{-1}$ & $3\times 10^{-6}$ \\
aluminum membrane \cite{Teufel2011} & $2.7$ & $48\,$pg & $1.1 \times 10^7$ & $3.3 \times 10^{5}$ & $0.015$ & $1.1 \times 10^7$ & $1\times 10^{-5}$ & $2\times 10^{-7}$
 \end{tabular}
\end{center}
\caption{CSL sensitivities for a selection of  optomechanical setups with density $\varrho$, mass $m$, mechanical frequency $\Omega$, quality factor $Q$, and temperature $T$.
The thermal sensitivity $\Lambda_{T}$ and the measurement-induced sensitivity  $\Lambda_{\rm SQL}$ (at measurement frequency $\omega$) are given in Eqs.~(\ref{eq:thBound}) and (\ref{eq:mBound}), respectively.
For simplicity, all systems are assumed to be center-of-mass oscillators, ignoring the specific mode profiles in \cite{Teufel2011,Purdy2013,Matsumoto2013}; the hypothetical setup is extrapolated from \cite{Matsumoto2013}, using a disc radius of $0.4\,$mm and thickness of $0.1\,$mm. A significant test of CSL requires 
$\Lambda_{T,\rm SQL} < 10^{-8 \pm 2}\,$Hz.}
\label{tab:CSL}
\end{table*}

\paragraph*{Bounds for the DP gravitational collapse model.} Compared to CSL, the Di\'osi-Penrose model exhibits a different mass dependence of the collapse effect, which is related to the gravitational self-energy of the mass distribution $\varrho(\vr)$ \cite{Diosi1987,Penrose1996}. The DP counterpart of the CSL diffusion rate reads as
\begin{align}
 D_\DP &= \frac{G \hbar}{2\pi^2} \int \diff^3 k \, \frac{k_x^2}{\vk^2} \left| \tilde{\varrho} (\vk) \right|^2,  \label{eq:DDP}
\end{align}
with $G$ the gravitational constant. Although this equation involves only natural constants,  one must introduce a 
blurring parameter $\sigma_\DP>0$ 
to account for the fact that the DP collapse effect diverges for point masses \cite{Ghirardi1990a,Bassi2003}. 
This implies that the DP effect depends not only on the macroscopic geometry of a given piece of matter, but is also highly sensitive to its microscopic 
lattice structure. That is, a cantilever can neither be assumed a homogeneous nor a rigid body. 

We model each nucleus in the crystal lattice as a Gaussian mass distribution of width $\sigma_\DP$, 
which determines the value of the DP diffusion rate (\ref{eq:DDP}) and can therefore be probed with optomechanical systems. 
For simplicity, we focus on monoatomic cubic lattices with lattice constant $a\gg \sigma_\DP$, where the DP diffusion rate reduces to the mass-proportional expression \cite{Supplement}
\begin{align}
  D_\DP &\approx \frac{G\hbar}{6\sqrt{\pi}} \left( \frac{a}{\sigma_\DP} \right)^3 \varrho m . \label{eq:DDPlattice}
\end{align}
The greatest detectable blurring parameter $\Sigma_\DP$ in the presence of thermal and measurement-induced noise is then mass-\emph{independent},
\begin{align}
  \sigma_\DP < \Sigma_\DP \equiv \left[ \frac{G\hbar \varrho}{6\sqrt{\pi}\left( \hbar\omega^2 + 2\gamma k_B T \right)} \right]^{1/3} a. \label{eq:DPbound}
\end{align}
Hence, also in this case one does not gain in sensitivity by increasing the oscillator's mass.

\paragraph*{Conclusion.}

We identified the generic sensitivity requirements for detecting stochastic collapse models in optomechanical setups. Since the predicted diffusion competes with inevitable thermal and measurement-induced noise, a high degree of experimental precision and control is crucial. A heavy oscillator, on the other hand, does not improve the sensitivity, even though the collapse-induced diffusion amplifies with mass: The contributions of thermal, back-action and shot noise grow in proportion to mass, whereas the growth of collapse-induced noise is generally weaker.

Our results show that one should rather aim for high quality factors, low and independently measured temperatures, and low oscillation frequencies. Precision experiments with micromechanical oscillators are insofar complementary to interferometric tests with delocalized nanoparticles \cite{Romero-Isart2011b,Nimmrichter2011a,Hornberger2012_RMP,Bateman2013}, where mass always matters.

%\bibliographystyle{apsrev4-1}
%\bibliography{library}

\begin{thebibliography}{46}%
\makeatletter
\providecommand \@ifxundefined [1]{%
 \@ifx{#1\undefined}
}%
\providecommand \@ifnum [1]{%
 \ifnum #1\expandafter \@firstoftwo
 \else \expandafter \@secondoftwo
 \fi
}%
\providecommand \@ifx [1]{%
 \ifx #1\expandafter \@firstoftwo
 \else \expandafter \@secondoftwo
 \fi
}%
\providecommand \natexlab [1]{#1}%
\providecommand \enquote  [1]{``#1''}%
\providecommand \bibnamefont  [1]{#1}%
\providecommand \bibfnamefont [1]{#1}%
\providecommand \citenamefont [1]{#1}%
\providecommand \href@noop [0]{\@secondoftwo}%
\providecommand \href [0]{\begingroup \@sanitize@url \@href}%
\providecommand \@href[1]{\@@startlink{#1}\@@href}%
\providecommand \@@href[1]{\endgroup#1\@@endlink}%
\providecommand \@sanitize@url [0]{\catcode `\\12\catcode `\$12\catcode
  `\&12\catcode `\#12\catcode `\^12\catcode `\_12\catcode `\%12\relax}%
\providecommand \@@startlink[1]{}%
\providecommand \@@endlink[0]{}%
\providecommand \url  [0]{\begingroup\@sanitize@url \@url }%
\providecommand \@url [1]{\endgroup\@href {#1}{\urlprefix }}%
\providecommand \urlprefix  [0]{URL }%
\providecommand \Eprint [0]{\href }%
\providecommand \doibase [0]{http://dx.doi.org/}%
\providecommand \selectlanguage [0]{\@gobble}%
\providecommand \bibinfo  [0]{\@secondoftwo}%
\providecommand \bibfield  [0]{\@secondoftwo}%
\providecommand \translation [1]{[#1]}%
\providecommand \BibitemOpen [0]{}%
\providecommand \bibitemStop [0]{}%
\providecommand \bibitemNoStop [0]{.\EOS\space}%
\providecommand \EOS [0]{\spacefactor3000\relax}%
\providecommand \BibitemShut  [1]{\csname bibitem#1\endcsname}%
\let\auto@bib@innerbib\@empty
%</preamble>
\bibitem [{\citenamefont {Hammerer}\ \emph {et~al.}(2010)\citenamefont
  {Hammerer}, \citenamefont {S\"{o}rensen},\ and\ \citenamefont
  {Polzik}}]{Hammerer2010}%
  \BibitemOpen
  \bibfield  {author} {\bibinfo {author} {\bibfnamefont {K.}~\bibnamefont
  {Hammerer}}, \bibinfo {author} {\bibfnamefont {A.~S.}\ \bibnamefont
  {S\"{o}rensen}}, \ and\ \bibinfo {author} {\bibfnamefont {E.~S.}\
  \bibnamefont {Polzik}},\ }\href@noop {} {\bibfield  {journal} {\bibinfo
  {journal} {Rev. Mod. Phys.}\ }\textbf {\bibinfo {volume} {82}},\
  \bibinfo {pages} {1041} (\bibinfo {year} {2010})}\BibitemShut {NoStop}%
\bibitem [{\citenamefont {De~Martini}\ and\ \citenamefont
  {Sciarrino}(2012)}]{RevModPhys.84.1765}%
  \BibitemOpen
  \bibfield  {author} {\bibinfo {author} {\bibfnamefont {F.}~\bibnamefont
  {De~Martini}}\ and\ \bibinfo {author} {\bibfnamefont {F.}~\bibnamefont
  {Sciarrino}},\ }\href {\doibase 10.1103/RevModPhys.84.1765} {\bibfield
  {journal} {\bibinfo  {journal} {Rev. Mod. Phys.}\ }\textbf {\bibinfo {volume}
  {84}},\ \bibinfo {pages} {1765} (\bibinfo {year} {2012})}\BibitemShut
  {NoStop}%
\bibitem [{\citenamefont {Pan}\ \emph {et~al.}(2012)\citenamefont {Pan},
  \citenamefont {Chen}, \citenamefont {Lu}, \citenamefont {Weinfurter},
  \citenamefont {Zeilinger},\ and\ \citenamefont
  {{Z}ukowski}}]{RevModPhys.84.777}%
  \BibitemOpen
  \bibfield  {author} {\bibinfo {author} {\bibfnamefont {J.-W.}\ \bibnamefont
  {Pan}}, \bibinfo {author} {\bibfnamefont {Z.-B.}\ \bibnamefont {Chen}},
  \bibinfo {author} {\bibfnamefont {C.-Y.}\ \bibnamefont {Lu}}, \bibinfo
  {author} {\bibfnamefont {H.}~\bibnamefont {Weinfurter}}, \bibinfo {author}
  {\bibfnamefont {A.}~\bibnamefont {Zeilinger}}, \ and\ \bibinfo {author}
  {\bibfnamefont {M.}~\bibnamefont {{Z}ukowski}},\ }\href {\doibase
  10.1103/RevModPhys.84.777} {\bibfield  {journal} {\bibinfo  {journal} {Rev.
  Mod. Phys.}\ }\textbf {\bibinfo {volume} {84}},\ \bibinfo {pages} {777}
  (\bibinfo {year} {2012})}\BibitemShut {NoStop}%
\bibitem [{\citenamefont {Arndt}\ and\ \citenamefont
  {Hornberger}(2014)}]{arndt_testing_2014}%
  \BibitemOpen
  \bibfield  {author} {\bibinfo {author} {\bibfnamefont {M.}~\bibnamefont
  {Arndt}}\ and\ \bibinfo {author} {\bibfnamefont {K.}~\bibnamefont
  {Hornberger}},\ }\href {\doibase 10.1038/nphys2863} {\bibfield  {journal}
  {\bibinfo  {journal} {Nat. Phys.}\ }\textbf {\bibinfo {volume} {10}},\ \bibinfo
  {pages} {271} (\bibinfo {year} {2014})}\BibitemShut {NoStop}%
\bibitem [{\citenamefont {Joos}\ \emph {et~al.}(2003)\citenamefont {Joos},
  \citenamefont {Zeh}, \citenamefont {Kiefer}, \citenamefont {Giulini},
  \citenamefont {Kupsch},\ and\ \citenamefont
  {Stamatescu}}]{joos_decoherence_2003}%
  \BibitemOpen
  \bibfield  {author} {\bibinfo {author} {\bibfnamefont {E.}~\bibnamefont
  {Joos}}, \bibinfo {author} {\bibfnamefont {H.~D.}\ \bibnamefont {Zeh}},
  \bibinfo {author} {\bibfnamefont {C.}~\bibnamefont {Kiefer}}, \bibinfo
  {author} {\bibfnamefont {D.}~\bibnamefont {Giulini}}, \bibinfo {author}
  {\bibfnamefont {J.}~\bibnamefont {Kupsch}}, \ and\ \bibinfo {author}
  {\bibfnamefont {I.-O.}\ \bibnamefont {Stamatescu}},\ }\href@noop {} {\emph
  {\bibinfo {title} {Decoherence and the Appearance of a Classical World in
  Quantum Theory}}}\ (\bibinfo  {publisher} {Springer},\ \bibinfo {address}
  {Berlin},\ \bibinfo {year} {2003})\BibitemShut {NoStop}%
\bibitem [{\citenamefont {Zurek}(2003)}]{Zurek2003}%
  \BibitemOpen
  \bibfield  {author} {\bibinfo {author} {\bibfnamefont {W.}~\bibnamefont
  {Zurek}},\ }\href {http://rmp.aps.org/abstract/RMP/v75/i3/p715\_1} {\bibfield
   {journal} {\bibinfo  {journal} {Rev. Mod. Phys.}\ }\textbf
  {\bibinfo {volume} {75}},\ \bibinfo {pages} {715} (\bibinfo {year}
  {2003})}\BibitemShut {NoStop}%
\bibitem [{\citenamefont {Schlosshauer}(2005)}]{Schlosshauer2005}%
  \BibitemOpen
  \bibfield  {author} {\bibinfo {author} {\bibfnamefont {M.}~\bibnamefont
  {Schlosshauer}},\ }\href {\doibase 10.1103/RevModPhys.76.1267} {\bibfield
  {journal} {\bibinfo  {journal} {Rev. Mod. Phys.}\ }\textbf
  {\bibinfo {volume} {76}},\ \bibinfo {pages} {1267} (\bibinfo {year}
  {2005})},\ \Eprint {http://arxiv.org/abs/0312059v4} {0312059v4} \BibitemShut
  {NoStop}%
\bibitem [{\citenamefont {Leggett}(2002)}]{Leggett2002}%
  \BibitemOpen
  \bibfield  {author} {\bibinfo {author} {\bibfnamefont {A.~J.}\ \bibnamefont
  {Leggett}},\ }\href@noop {} {\bibfield  {journal} {\bibinfo  {journal}
  {J. Phys. Cond. Mat.}\ }\textbf {\bibinfo {volume} {14}},\
  \bibinfo {pages} {R415} (\bibinfo {year} {2002})}\BibitemShut {NoStop}%
\bibitem [{\citenamefont {Bassi}\ and\ \citenamefont
  {Ghirardi}(2003)}]{Bassi2003}%
  \BibitemOpen
  \bibfield  {author} {\bibinfo {author} {\bibfnamefont {A.}~\bibnamefont
  {Bassi}}\ and\ \bibinfo {author} {\bibfnamefont {G.}~\bibnamefont
  {Ghirardi}},\ }\href@noop {} {\bibfield  {journal} {\bibinfo  {journal}
  {Phys. Rep.}\ }\textbf {\bibinfo {volume} {379}},\ \bibinfo {pages}
  {257} (\bibinfo {year} {2003})}\BibitemShut {NoStop}%
\bibitem [{\citenamefont {Ghirardi}\ \emph
  {et~al.}(1990{\natexlab{a}})\citenamefont {Ghirardi}, \citenamefont
  {Pearle},\ and\ \citenamefont {Rimini}}]{Ghirardi1990b}%
  \BibitemOpen
  \bibfield  {author} {\bibinfo {author} {\bibfnamefont {G.~C.}\ \bibnamefont
  {Ghirardi}}, \bibinfo {author} {\bibfnamefont {P.}~\bibnamefont {Pearle}}, \
  and\ \bibinfo {author} {\bibfnamefont {A.}~\bibnamefont {Rimini}},\
  }\href@noop {} {\bibfield  {journal} {\bibinfo  {journal} {Phys. Rev.
  A}\ }\textbf {\bibinfo {volume} {42}},\ \bibinfo {pages} {78} (\bibinfo
  {year} {1990}{\natexlab{a}})}\BibitemShut {NoStop}%
\bibitem [{\citenamefont {Bassi}\ \emph {et~al.}(2013)\citenamefont {Bassi},
  \citenamefont {Lochan}, \citenamefont {Satin}, \citenamefont {Singh},\ and\
  \citenamefont {Ulbricht}}]{Bassi2013}%
  \BibitemOpen
  \bibfield  {author} {\bibinfo {author} {\bibfnamefont {A.}~\bibnamefont
  {Bassi}}, \bibinfo {author} {\bibfnamefont {K.}~\bibnamefont {Lochan}},
  \bibinfo {author} {\bibfnamefont {S.}~\bibnamefont {Satin}}, \bibinfo
  {author} {\bibfnamefont {T.~P.}\ \bibnamefont {Singh}}, \ and\ \bibinfo
  {author} {\bibfnamefont {H.}~\bibnamefont {Ulbricht}},\ }\href {\doibase
  10.1103/RevModPhys.85.471} {\bibfield  {journal} {\bibinfo  {journal}
  {Rev. Mod. Phys.}\ }\textbf {\bibinfo {volume} {85}},\ \bibinfo
  {pages} {471} (\bibinfo {year} {2013})}\BibitemShut {NoStop}%
\bibitem [{\citenamefont {Adler}(2004)}]{Adler2004}%
  \BibitemOpen
  \bibfield  {author} {\bibinfo {author} {\bibfnamefont {S.~L.}\ \bibnamefont
  {Adler}},\ }\href
  {http://www.amazon.com/Quantum-Theory-Emergent-Phenomenon-Statistical/dp/0521831946}
  {\emph {\bibinfo {title} {{Quantum Theory as an Emergent Phenomenon: The
  Statistical Mechanics of Matrix Models as the Precursor of Quantum Field
  Theory}}}}\ (\bibinfo  {publisher} {Cambridge University Press},\ \bibinfo
  {year} {2004})\BibitemShut {NoStop}%
\bibitem [{\citenamefont {Nimmrichter}\ and\ \citenamefont
  {Hornberger}(2013)}]{Nimmrichter2013}%
  \BibitemOpen
  \bibfield  {author} {\bibinfo {author} {\bibfnamefont {S.}~\bibnamefont
  {Nimmrichter}}\ and\ \bibinfo {author} {\bibfnamefont {K.}~\bibnamefont
  {Hornberger}},\ }\href {\doibase 10.1103/PhysRevLett.110.160403} {\bibfield
  {journal} {\bibinfo  {journal} {Phys. Rev. Lett.}\ }\textbf {\bibinfo
  {volume} {110}},\ \bibinfo {pages} {160403} (\bibinfo {year}
  {2013})}\BibitemShut {NoStop}%
\bibitem [{\citenamefont {Di\'{o}si}(1987)}]{Diosi1987}%
  \BibitemOpen
  \bibfield  {author} {\bibinfo {author} {\bibfnamefont {L.}~\bibnamefont
  {Di\'{o}si}},\ }\href {\doibase 10.1016/0375-9601(87)90681-5} {\bibfield
  {journal} {\bibinfo  {journal} {Phys. Lett. A}\ }\textbf {\bibinfo
  {volume} {120}},\ \bibinfo {pages} {377} (\bibinfo {year}
  {1987})}\BibitemShut {NoStop}%
\bibitem [{\citenamefont {Di\'{o}si}(1989)}]{Diosi1989}%
  \BibitemOpen
  \bibfield  {author} {\bibinfo {author} {\bibfnamefont {L.}~\bibnamefont
  {Di\'{o}si}},\ }\href {\doibase 10.1103/PhysRevA.40.1165} {\bibfield
  {journal} {\bibinfo  {journal} {Phys. Rev. A}\ }\textbf {\bibinfo
  {volume} {40}},\ \bibinfo {pages} {1165} (\bibinfo {year}
  {1989})}\BibitemShut {NoStop}%
\bibitem [{\citenamefont {Penrose}(1996)}]{Penrose1996}%
  \BibitemOpen
  \bibfield  {author} {\bibinfo {author} {\bibfnamefont {R.}~\bibnamefont
  {Penrose}},\ }\href {http://www.springerlink.com/index/k75046wh3668l654.pdf}
  {\bibfield  {journal} {\bibinfo  {journal} {Gen. Rel. Grav.}\ }\textbf {\bibinfo {volume} {28}},\ \bibinfo {pages} {581}
  (\bibinfo {year} {1996})}\BibitemShut {NoStop}%
\bibitem [{\citenamefont {Marshall}\ \emph {et~al.}(2003)\citenamefont
  {Marshall}, \citenamefont {Simon}, \citenamefont {Penrose},\ and\
  \citenamefont {Bouwmeester}}]{Marshall2003}%
  \BibitemOpen
  \bibfield  {author} {\bibinfo {author} {\bibfnamefont {W.}~\bibnamefont
  {Marshall}}, \bibinfo {author} {\bibfnamefont {C.}~\bibnamefont {Simon}},
  \bibinfo {author} {\bibfnamefont {R.}~\bibnamefont {Penrose}}, \ and\
  \bibinfo {author} {\bibfnamefont {D.}~\bibnamefont {Bouwmeester}},\ }\href
  {\doibase 10.1103/PhysRevLett.91.130401} {\bibfield  {journal} {\bibinfo
  {journal} {Phys. Rev. Lett.}\ }\textbf {\bibinfo {volume} {91}},\
  \bibinfo {pages} {130401} (\bibinfo {year} {2003})}\BibitemShut {NoStop}%
\bibitem [{\citenamefont {Romero-Isart}\ \emph {et~al.}(2011)\citenamefont
  {Romero-Isart}, \citenamefont {Pflanzer}, \citenamefont {Blaser},
  \citenamefont {Kaltenbaek}, \citenamefont {Kiesel}, \citenamefont
  {Aspelmeyer},\ and\ \citenamefont {Cirac}}]{Romero-Isart2011b}%
  \BibitemOpen
  \bibfield  {author} {\bibinfo {author} {\bibfnamefont {O.}~\bibnamefont
  {Romero-Isart}}, \bibinfo {author} {\bibfnamefont {A.~C.}\ \bibnamefont
  {Pflanzer}}, \bibinfo {author} {\bibfnamefont {F.}~\bibnamefont {Blaser}},
  \bibinfo {author} {\bibfnamefont {R.}~\bibnamefont {Kaltenbaek}}, \bibinfo
  {author} {\bibfnamefont {N.}~\bibnamefont {Kiesel}}, \bibinfo {author}
  {\bibfnamefont {M.}~\bibnamefont {Aspelmeyer}}, \ and\ \bibinfo {author}
  {\bibfnamefont {J.~I.}\ \bibnamefont {Cirac}},\ }\href@noop {} {\bibfield
  {journal} {\bibinfo  {journal} {Phys. Rev. Lett.}\ }\textbf {\bibinfo
  {volume} {107}},\ \bibinfo {pages} {20405} (\bibinfo {year}
  {2011})}\BibitemShut {NoStop}%
\bibitem [{\citenamefont {Nimmrichter}\ \emph {et~al.}(2011)\citenamefont
  {Nimmrichter}, \citenamefont {Hornberger}, \citenamefont {Haslinger},\ and\
  \citenamefont {Arndt}}]{Nimmrichter2011a}%
  \BibitemOpen
  \bibfield  {author} {\bibinfo {author} {\bibfnamefont {S.}~\bibnamefont
  {Nimmrichter}}, \bibinfo {author} {\bibfnamefont {K.}~\bibnamefont
  {Hornberger}}, \bibinfo {author} {\bibfnamefont {P.}~\bibnamefont
  {Haslinger}}, \ and\ \bibinfo {author} {\bibfnamefont {M.}~\bibnamefont
  {Arndt}},\ }\href@noop {} {\bibfield  {journal} {\bibinfo  {journal}
  {Phys. Rev. A}\ }\textbf {\bibinfo {volume} {83}},\ \bibinfo {pages}
  {43621} (\bibinfo {year} {2011})}\BibitemShut {NoStop}%
\bibitem [{\citenamefont {Hornberger}\ \emph {et~al.}(2012)\citenamefont
  {Hornberger}, \citenamefont {Gerlich}, \citenamefont {Haslinger},
  \citenamefont {Nimmrichter},\ and\ \citenamefont
  {Arndt}}]{Hornberger2012_RMP}%
  \BibitemOpen
  \bibfield  {author} {\bibinfo {author} {\bibfnamefont {K.}~\bibnamefont
  {Hornberger}}, \bibinfo {author} {\bibfnamefont {S.}~\bibnamefont {Gerlich}},
  \bibinfo {author} {\bibfnamefont {P.}~\bibnamefont {Haslinger}}, \bibinfo
  {author} {\bibfnamefont {S.}~\bibnamefont {Nimmrichter}}, \ and\ \bibinfo
  {author} {\bibfnamefont {M.}~\bibnamefont {Arndt}},\ }\href@noop {}
  {\bibfield  {journal} {\bibinfo  {journal} {Rev. Mod. Phys.}\
  }\textbf {\bibinfo {volume} {84}},\ \bibinfo {pages} {157} (\bibinfo {year}
  {2012})}\BibitemShut {NoStop}%
\bibitem [{\citenamefont {Bateman}\ \emph {et~al.}(2013)\citenamefont
  {Bateman}, \citenamefont {Nimmrichter}, \citenamefont {Hornberger},\ and\
  \citenamefont {Ulbricht}}]{Bateman2013}%
  \BibitemOpen
  \bibfield  {author} {\bibinfo {author} {\bibfnamefont {J.}~\bibnamefont
  {Bateman}}, \bibinfo {author} {\bibfnamefont {S.}~\bibnamefont
  {Nimmrichter}}, \bibinfo {author} {\bibfnamefont {K.}~\bibnamefont
  {Hornberger}}, \ and\ \bibinfo {author} {\bibfnamefont {H.}~\bibnamefont
  {Ulbricht}},\ }\href {http://arxiv.org/abs/1312.0500} {\bibfield  {journal}
  {\bibinfo  {journal} {arXiv:1312.0500 [quant-ph]}\ } (\bibinfo
  {year} {2013})}
  \BibitemShut {NoStop}%
\bibitem [{\citenamefont {Bahrami}\ \emph {et~al.}(2014)\citenamefont
  {Bahrami}, \citenamefont {Paternostro}, \citenamefont {Bassi},\ and\
  \citenamefont {Ulbricht}}]{Bahrami2014}%
  \BibitemOpen
  \bibfield  {author} {\bibinfo {author} {\bibfnamefont {M.}~\bibnamefont
  {Bahrami}}, \bibinfo {author} {\bibfnamefont {M.}~\bibnamefont
  {Paternostro}}, \bibinfo {author} {\bibfnamefont {A.}~\bibnamefont {Bassi}},
  \ and\ \bibinfo {author} {\bibfnamefont {H.}~\bibnamefont {Ulbricht}},\
  }\href {\doibase 10.1103/PhysRevLett.112.210404} {\bibfield  {journal}
  {\bibinfo  {journal} {Phys. Rev. Lett.}\ }\textbf {\bibinfo {volume} {112}},\
  \bibinfo {pages} {210404} (\bibinfo {year} {2014})}\BibitemShut {NoStop}%
\bibitem [{\citenamefont {Clerk}\ \emph {et~al.}(2010)\citenamefont {Clerk},
  \citenamefont {Devoret}, \citenamefont {Girvin}, \citenamefont {Marquardt},\
  and\ \citenamefont {Schoelkopf}}]{Clerk2010a}%
  \BibitemOpen
  \bibfield  {author} {\bibinfo {author} {\bibfnamefont {A.~A.}\ \bibnamefont
  {Clerk}}, \bibinfo {author} {\bibfnamefont {M.~H.}\ \bibnamefont {Devoret}},
  \bibinfo {author} {\bibfnamefont {S.~M.}\ \bibnamefont {Girvin}}, \bibinfo
  {author} {\bibfnamefont {F.}~\bibnamefont {Marquardt}}, \ and\ \bibinfo
  {author} {\bibfnamefont {R.~J.}\ \bibnamefont {Schoelkopf}},\ }\href
  {http://link.aps.org/doi/10.1103/RevModPhys.82.1155} {\bibfield  {journal}
  {\bibinfo  {journal} {Rev. Mod. Phys.}\ }\textbf {\bibinfo {volume}
  {82}},\ \bibinfo {pages} {1155} (\bibinfo {year} {2010})}\BibitemShut
  {NoStop}%
\bibitem [{\citenamefont {Aspelmeyer}\ \emph {et~al.}(2013)\citenamefont
  {Aspelmeyer}, \citenamefont {Kippenberg},\ and\ \citenamefont
  {Marquardt}}]{Aspelmeyer2013}%
  \BibitemOpen
  \bibfield  {author} {\bibinfo {author} {\bibfnamefont {M.}~\bibnamefont
  {Aspelmeyer}}, \bibinfo {author} {\bibfnamefont {T.~J.}\ \bibnamefont
  {Kippenberg}}, \ and\ \bibinfo {author} {\bibfnamefont {F.}~\bibnamefont
  {Marquardt}},\ }\href {http://arxiv.org/abs/1303.0733} {\bibfield  {journal}
  {\bibinfo  {journal} {arXiv:1303.0733 [cond-mat.mes-hall]}\ } (\bibinfo
  {year} {2013})}
  \BibitemShut {NoStop}%
\bibitem [{\citenamefont {Danilishin}\ and\ \citenamefont
  {Khalili}(2012)}]{Danilishin2012}%
  \BibitemOpen
  \bibfield  {author} {\bibinfo {author} {\bibfnamefont {S.~L.}\ \bibnamefont
  {Danilishin}}\ and\ \bibinfo {author} {\bibfnamefont {F.~Y.}\ \bibnamefont
  {Khalili}},\ }\href
  {http://relativity.livingreviews.org/Articles/lrr-2012-5/} {\bibfield
  {journal} {\bibinfo  {journal} {Living Rev. Relativity}\ }\textbf {\bibinfo
  {volume} {15}} (\bibinfo {year} {2012})}\BibitemShut {NoStop}%
\bibitem [{\citenamefont {Chen}(2013)}]{Chen2013}%
  \BibitemOpen
  \bibfield  {author} {\bibinfo {author} {\bibfnamefont {Y.}~\bibnamefont
  {Chen}},\ }\href {\doibase 10.1088/0953-4075/46/10/104001} {\bibfield
  {journal} {\bibinfo  {journal} {J. Phys. B: Atomic, Molecular and
  Optical Physics}\ }\textbf {\bibinfo {volume} {46}},\ \bibinfo {pages}
  {104001} (\bibinfo {year} {2013})}\BibitemShut {NoStop}%
\bibitem [{\citenamefont {Murch}\ \emph {et~al.}(2008)\citenamefont {Murch},
  \citenamefont {Moore}, \citenamefont {Gupta},\ and\ \citenamefont
  {Stamper-Kurn}}]{Murch2008}%
  \BibitemOpen
  \bibfield  {author} {\bibinfo {author} {\bibfnamefont {K.~W.}\ \bibnamefont
  {Murch}}, \bibinfo {author} {\bibfnamefont {K.~L.}\ \bibnamefont {Moore}},
  \bibinfo {author} {\bibfnamefont {S.}~\bibnamefont {Gupta}}, \ and\ \bibinfo
  {author} {\bibfnamefont {D.~M.}\ \bibnamefont {Stamper-Kurn}},\ }\href
  {\doibase 10.1038/nphys965} {\bibfield  {journal} {\bibinfo  {journal}
  {Nat. Phys.}\ }\textbf {\bibinfo {volume} {4}},\ \bibinfo {pages} {561}
  (\bibinfo {year} {2008})}\BibitemShut {NoStop}%
\bibitem [{\citenamefont {Safavi-Naeini}\ \emph {et~al.}(2012)\citenamefont
  {Safavi-Naeini}, \citenamefont {Chan}, \citenamefont {Hill}, \citenamefont
  {Alegre}, \citenamefont {Krause},\ and\ \citenamefont
  {Painter}}]{Safavi-Naeini2012}%
  \BibitemOpen
  \bibfield  {author} {\bibinfo {author} {\bibfnamefont {A.~H.}\ \bibnamefont
  {Safavi-Naeini}}, \bibinfo {author} {\bibfnamefont {J.}~\bibnamefont {Chan}},
  \bibinfo {author} {\bibfnamefont {J.~T.}\ \bibnamefont {Hill}}, \bibinfo
  {author} {\bibfnamefont {T.~P.~M.}\ \bibnamefont {Alegre}}, \bibinfo {author}
  {\bibfnamefont {A.}~\bibnamefont {Krause}}, \ and\ \bibinfo {author}
  {\bibfnamefont {O.}~\bibnamefont {Painter}},\ }\href {\doibase
  10.1103/PhysRevLett.108.033602} {\bibfield  {journal} {\bibinfo  {journal}
  {Phys. Rev. Lett.}\ }\textbf {\bibinfo {volume} {108}},\ \bibinfo {pages}
  {033602} (\bibinfo {year} {2012})}\BibitemShut {NoStop}%
\bibitem [{\citenamefont {Purdy}\ \emph {et~al.}(2013)\citenamefont {Purdy},
  \citenamefont {Peterson},\ and\ \citenamefont {Regal}}]{Purdy2013}%
  \BibitemOpen
  \bibfield  {author} {\bibinfo {author} {\bibfnamefont {T.~P.}\ \bibnamefont
  {Purdy}}, \bibinfo {author} {\bibfnamefont {R.~W.}\ \bibnamefont {Peterson}},
  \ and\ \bibinfo {author} {\bibfnamefont {C.~A.}\ \bibnamefont {Regal}},\
  }\href {\doibase 10.1126/science.1231282} {\bibfield  {journal} {\bibinfo
  {journal} {Science}\ }\textbf {\bibinfo {volume} {339}},\
  \bibinfo {pages} {801} (\bibinfo {year} {2013})}\BibitemShut {NoStop}%
\bibitem [{\citenamefont {Matsumoto}\ \emph {et~al.}(2013)\citenamefont
  {Matsumoto}, \citenamefont {Michimura}, \citenamefont {Hayase}, \citenamefont
  {Aso},\ and\ \citenamefont {Tsubono}}]{Matsumoto2013}%
  \BibitemOpen
  \bibfield  {author} {\bibinfo {author} {\bibfnamefont {N.}~\bibnamefont
  {Matsumoto}}, \bibinfo {author} {\bibfnamefont {Y.}~\bibnamefont
  {Michimura}}, \bibinfo {author} {\bibfnamefont {G.}~\bibnamefont {Hayase}},
  \bibinfo {author} {\bibfnamefont {Y.}~\bibnamefont {Aso}}, \ and\ \bibinfo
  {author} {\bibfnamefont {K.}~\bibnamefont {Tsubono}},\ }\href
  {http://arxiv.org/abs/1312.5031} {\  (\bibinfo {year} {2013})},\ \Eprint
  {http://arxiv.org/abs/1312.5031} {arXiv:1312.5031} \BibitemShut {NoStop}%
\bibitem [{\citenamefont {Caves}\ \emph {et~al.}(1980)\citenamefont {Caves},
  \citenamefont {Thorne}, \citenamefont {Drever}, \citenamefont {Sandberg},\
  and\ \citenamefont {Zimmermann}}]{Caves1980a}%
  \BibitemOpen
  \bibfield  {author} {\bibinfo {author} {\bibfnamefont {C.}~\bibnamefont
  {Caves}}, \bibinfo {author} {\bibfnamefont {K.}~\bibnamefont {Thorne}},
  \bibinfo {author} {\bibfnamefont {R.}~\bibnamefont {Drever}}, \bibinfo
  {author} {\bibfnamefont {V.}~\bibnamefont {Sandberg}}, \ and\ \bibinfo
  {author} {\bibfnamefont {M.}~\bibnamefont {Zimmermann}},\ }\href
  {http://link.aps.org/doi/10.1103/RevModPhys.52.341} {\bibfield  {journal}
  {\bibinfo  {journal} {Rev. Mod. Phys.}\ }\textbf {\bibinfo {volume}
  {52}},\ \bibinfo {pages} {341} (\bibinfo {year} {1980})}\BibitemShut
  {NoStop}%
\bibitem [{\citenamefont {Caves}(1981)}]{Caves1981}%
  \BibitemOpen
  \bibfield  {author} {\bibinfo {author} {\bibfnamefont {C.}~\bibnamefont
  {Caves}},\ }\href {\doibase 10.1103/PhysRevD.23.1693} {\bibfield  {journal}
  {\bibinfo  {journal} {Phys. Rev. D}\ }\textbf {\bibinfo {volume} {23}},\
  \bibinfo {pages} {1693} (\bibinfo {year} {1981})}\BibitemShut {NoStop}%
\bibitem [{\citenamefont {M\"uller-Ebhardt}\ \emph {et~al.}(2008)\citenamefont
  {M\"uller-Ebhardt}, \citenamefont {Rehbein}, \citenamefont {Schnabel},
  \citenamefont {Danzmann},\ and\ \citenamefont
  {Chen}}]{PhysRevLett.100.013601}%
  \BibitemOpen
  \bibfield  {author} {\bibinfo {author} {\bibfnamefont {H.}~\bibnamefont
  {M\"uller-Ebhardt}}, \bibinfo {author} {\bibfnamefont {H.}~\bibnamefont
  {Rehbein}}, \bibinfo {author} {\bibfnamefont {R.}~\bibnamefont {Schnabel}},
  \bibinfo {author} {\bibfnamefont {K.}~\bibnamefont {Danzmann}}, \ and\
  \bibinfo {author} {\bibfnamefont {Y.}~\bibnamefont {Chen}},\ }\href {\doibase
  10.1103/PhysRevLett.100.013601} {\bibfield  {journal} {\bibinfo  {journal}
  {Phys. Rev. Lett.}\ }\textbf {\bibinfo {volume} {100}},\ \bibinfo {pages}
  {013601} (\bibinfo {year} {2008})}\BibitemShut {NoStop}%
\bibitem [{\citenamefont {Miao}\ \emph {et~al.}(2010)\citenamefont {Miao},
  \citenamefont {Danilishin}, \citenamefont {M\"uller-Ebhardt}, \citenamefont
  {Rehbein}, \citenamefont {Somiya},\ and\ \citenamefont
  {Chen}}]{PhysRevA.81.012114}%
  \BibitemOpen
  \bibfield  {author} {\bibinfo {author} {\bibfnamefont {H.}~\bibnamefont
  {Miao}}, \bibinfo {author} {\bibfnamefont {S.}~\bibnamefont {Danilishin}},
  \bibinfo {author} {\bibfnamefont {H.}~\bibnamefont {M\"uller-Ebhardt}},
  \bibinfo {author} {\bibfnamefont {H.}~\bibnamefont {Rehbein}}, \bibinfo
  {author} {\bibfnamefont {K.}~\bibnamefont {Somiya}}, \ and\ \bibinfo {author}
  {\bibfnamefont {Y.}~\bibnamefont {Chen}},\ }\href {\doibase
  10.1103/PhysRevA.81.012114} {\bibfield  {journal} {\bibinfo  {journal} {Phys.
  Rev. A}\ }\textbf {\bibinfo {volume} {81}},\ \bibinfo {pages} {012114}
  (\bibinfo {year} {2010})}\BibitemShut {NoStop}%
\bibitem [{\citenamefont {Bassi}\ \emph {et~al.}(2005)\citenamefont {Bassi},
  \citenamefont {Ippoliti},\ and\ \citenamefont {Adler}}]{Bassi2005}%
  \BibitemOpen
  \bibfield  {author} {\bibinfo {author} {\bibfnamefont {A.}~\bibnamefont
  {Bassi}}, \bibinfo {author} {\bibfnamefont {E.}~\bibnamefont {Ippoliti}}, \
  and\ \bibinfo {author} {\bibfnamefont {S.}~\bibnamefont {Adler}},\
  }\href@noop {} {\bibfield  {journal} {\bibinfo  {journal} {Phys. Rev. Lett.}\ }\textbf {\bibinfo {volume} {94}},\ \bibinfo {pages} {30401}
  (\bibinfo {year} {2005})}\BibitemShut {NoStop}%
\bibitem [{\citenamefont {Fu}(1997)}]{Fu1997}%
  \BibitemOpen
  \bibfield  {author} {\bibinfo {author} {\bibfnamefont {Q.}~\bibnamefont
  {Fu}},\ }\href {\doibase 10.1103/PhysRevA.56.1806} {\bibfield  {journal}
  {\bibinfo  {journal} {Phys. Rev. A}\ }\textbf {\bibinfo {volume} {56}},\
  \bibinfo {pages} {1806} (\bibinfo {year} {1997})}\BibitemShut {NoStop}%
\bibitem [{\citenamefont {Adler}\ and\ \citenamefont
  {Ramazanoğlu}(2007)}]{Adler2007a}%
  \BibitemOpen
  \bibfield  {author} {\bibinfo {author} {\bibfnamefont {S.~L.}\ \bibnamefont
  {Adler}}\ and\ \bibinfo {author} {\bibfnamefont {F.~M.}\ \bibnamefont
  {Ramazanoğlu}},\ }\href {\doibase 10.1088/1751-8113/40/44/017} {\bibfield
  {journal} {\bibinfo  {journal} {J. Phys. A: Math. Theor.}\ }\textbf {\bibinfo
  {volume} {40}},\ \bibinfo {pages} {13395} (\bibinfo {year}
  {2007})}\BibitemShut {NoStop}%
\bibitem [{\citenamefont {Feldmann}\ and\ \citenamefont
  {Tumulka}(2012)}]{Feldmann2012}%
  \BibitemOpen
  \bibfield  {author} {\bibinfo {author} {\bibfnamefont {W.}~\bibnamefont
  {Feldmann}}\ and\ \bibinfo {author} {\bibfnamefont {R.}~\bibnamefont
  {Tumulka}},\ }\href {\doibase 10.1088/1751-8113/45/6/065304} {\bibfield
  {journal} {\bibinfo  {journal} {J. Phys. A: Math. Theor.}\ }\textbf {\bibinfo {volume} {45}},\ \bibinfo {pages} {065304}
  (\bibinfo {year} {2012})}\BibitemShut {NoStop}%
\bibitem [{\citenamefont {Adler}(2007)}]{Adler2007}%
  \BibitemOpen
  \bibfield  {author} {\bibinfo {author} {\bibfnamefont {S.~L.}\ \bibnamefont
  {Adler}},\ }\href@noop {} {\bibfield  {journal} {\bibinfo  {journal} {J. Phys. A: Math. Theor.}\ }\textbf {\bibinfo {volume}
  {40}},\ \bibinfo {pages} {2935} (\bibinfo {year} {2007})}\BibitemShut
  {NoStop}%
\bibitem [{\citenamefont {Chang}\ \emph {et~al.}(2010)\citenamefont {Chang},
  \citenamefont {Regal}, \citenamefont {Papp}, \citenamefont {Wilson},
  \citenamefont {Ye}, \citenamefont {Painter}, \citenamefont {Kimble},\ and\
  \citenamefont {Zoller}}]{Chang2010}%
  \BibitemOpen
  \bibfield  {author} {\bibinfo {author} {\bibfnamefont {D.~E.}\ \bibnamefont
  {Chang}}, \bibinfo {author} {\bibfnamefont {C.~A.}\ \bibnamefont {Regal}},
  \bibinfo {author} {\bibfnamefont {S.~B.}\ \bibnamefont {Papp}}, \bibinfo
  {author} {\bibfnamefont {D.~J.}\ \bibnamefont {Wilson}}, \bibinfo {author}
  {\bibfnamefont {J.}~\bibnamefont {Ye}}, \bibinfo {author} {\bibfnamefont
  {O.}~\bibnamefont {Painter}}, \bibinfo {author} {\bibfnamefont {H.~J.}\
  \bibnamefont {Kimble}}, \ and\ \bibinfo {author} {\bibfnamefont
  {P.}~\bibnamefont {Zoller}},\ }\href {\doibase 10.1073/pnas.0912969107}
  {\bibfield  {journal} {\bibinfo  {journal} {Proc. Natl. Acad. Sci.}\ }\textbf {\bibinfo {volume} {107}},\ \bibinfo {pages}
  {1005} (\bibinfo {year} {2010})}
  \BibitemShut {NoStop}%
\bibitem [{\citenamefont {Romero-Isart}\ \emph {et~al.}(2010)\citenamefont
  {Romero-Isart}, \citenamefont {Juan}, \citenamefont {Quidant},\ and\
  \citenamefont {Cirac}}]{Romero-Isart2010}%
  \BibitemOpen
  \bibfield  {author} {\bibinfo {author} {\bibfnamefont {O.}~\bibnamefont
  {Romero-Isart}}, \bibinfo {author} {\bibfnamefont {M.~L.}\ \bibnamefont
  {Juan}}, \bibinfo {author} {\bibfnamefont {R.}~\bibnamefont {Quidant}}, \
  and\ \bibinfo {author} {\bibfnamefont {J.~I.}\ \bibnamefont {Cirac}},\ }\href
  {\doibase 10.1088/1367-2630/12/3/033015} {\bibfield  {journal} {\bibinfo
  {journal} {New J. Phys.}\ }\textbf {\bibinfo {volume} {12}},\
  \bibinfo {pages} {033015} (\bibinfo {year} {2010})}\BibitemShut
  {NoStop}%
\bibitem [{\citenamefont {Thompson}\ \emph {et~al.}(2008)\citenamefont
  {Thompson}, \citenamefont {Zwickl}, \citenamefont {Jayich}, \citenamefont
  {Marquardt}, \citenamefont {Girvin},\ and\ \citenamefont
  {Harris}}]{Thompson2008}%
  \BibitemOpen
  \bibfield  {author} {\bibinfo {author} {\bibfnamefont {J.~D.}\ \bibnamefont
  {Thompson}}, \bibinfo {author} {\bibfnamefont {B.~M.}\ \bibnamefont
  {Zwickl}}, \bibinfo {author} {\bibfnamefont {A.~M.}\ \bibnamefont {Jayich}},
  \bibinfo {author} {\bibfnamefont {F.}~\bibnamefont {Marquardt}}, \bibinfo
  {author} {\bibfnamefont {S.~M.}\ \bibnamefont {Girvin}}, \ and\ \bibinfo
  {author} {\bibfnamefont {J.~G.~E.}\ \bibnamefont {Harris}},\ }\href {\doibase
  10.1038/nature06715} {\bibfield  {journal} {\bibinfo  {journal} {Nature}\
  }\textbf {\bibinfo {volume} {452}},\ \bibinfo {pages} {72} (\bibinfo {year}
  {2008})}\BibitemShut {NoStop}%
\bibitem [{\citenamefont {Teufel}\ \emph {et~al.}(2011)\citenamefont {Teufel},
  \citenamefont {Donner}, \citenamefont {Li}, \citenamefont {Harlow},
  \citenamefont {Allman}, \citenamefont {Cicak}, \citenamefont {Sirois},
  \citenamefont {Whittaker}, \citenamefont {Lehnert},\ and\ \citenamefont
  {Simmonds}}]{Teufel2011}%
  \BibitemOpen
  \bibfield  {author} {\bibinfo {author} {\bibfnamefont {J.~D.}\ \bibnamefont
  {Teufel}}, \bibinfo {author} {\bibfnamefont {T.}~\bibnamefont {Donner}},
  \bibinfo {author} {\bibfnamefont {D.}~\bibnamefont {Li}}, \bibinfo {author}
  {\bibfnamefont {J.~W.}\ \bibnamefont {Harlow}}, \bibinfo {author}
  {\bibfnamefont {M.~S.}\ \bibnamefont {Allman}}, \bibinfo {author}
  {\bibfnamefont {K.}~\bibnamefont {Cicak}}, \bibinfo {author} {\bibfnamefont
  {A.~J.}\ \bibnamefont {Sirois}}, \bibinfo {author} {\bibfnamefont {J.~D.}\
  \bibnamefont {Whittaker}}, \bibinfo {author} {\bibfnamefont {K.~W.}\
  \bibnamefont {Lehnert}}, \ and\ \bibinfo {author} {\bibfnamefont {R.~W.}\
  \bibnamefont {Simmonds}},\ }\href {\doibase 10.1038/nature10261} {\bibfield
  {journal} {\bibinfo  {journal} {Nature}\ }\textbf {\bibinfo {volume} {475}},\
  \bibinfo {pages} {359} (\bibinfo {year} {2011})}\BibitemShut {NoStop}%
\bibitem [{Sup()}]{Supplement}%
  \BibitemOpen
  \href@noop {} {}\bibinfo {note} {See Supplemental Material below for
  details on the collapse-induced diffusion rates.}\BibitemShut {Stop}%
\bibitem [{Not()}]{Note1}%
  \BibitemOpen
  \href@noop {} {}\bibinfo {note} {Although we focus on center-of-mass
  oscillations, the considered collapse models should imply a similar size
  dependence for other mechanical modes with smaller effective
  masses.}\BibitemShut {Stop}%
\bibitem [{\citenamefont {Ghirardi}\ \emph
  {et~al.}(1990{\natexlab{b}})\citenamefont {Ghirardi}, \citenamefont
  {Grassi},\ and\ \citenamefont {Rimini}}]{Ghirardi1990a}%
  \BibitemOpen
  \bibfield  {author} {\bibinfo {author} {\bibfnamefont {G.}~\bibnamefont
  {Ghirardi}}, \bibinfo {author} {\bibfnamefont {R.}~\bibnamefont {Grassi}}, \
  and\ \bibinfo {author} {\bibfnamefont {A.}~\bibnamefont {Rimini}},\ }\href
  {\doibase 10.1103/PhysRevA.42.1057} {\bibfield  {journal} {\bibinfo
  {journal} {Phys. Rev. A}\ }\textbf {\bibinfo {volume} {42}},\ \bibinfo
  {pages} {1057} (\bibinfo {year} {1990}{\natexlab{b}})}\BibitemShut {NoStop}%
\end{thebibliography}


%\clearpage
\appendix

\onecolumngrid

\section*{Supplemental Material}

\setcounter{equation}{0}
\renewcommand{\theequation}{S\arabic{equation}}

\section{Diffusion predicted by CSL and DP}

Here we derive explicitly the momentum diffusion rates $D_\CSL$ and $D_\DP$ predicted by the CSL and the DP model, which result in Equations (1) and (11) in the main text. 

The CSL master equation for a system of $N$ masses $m_n$ with position operators $\ovr_n$ reads in a first-quantization picture as \cite{Bassi2003,Bassi2013}
%\begin{align}
% \cL_\CSL = \frac{8\pi^{3/2}r^3_\CSL \lambda_\CSL}{{\rm amu}^2} \int \diff^3 s \left[ \om(\vs) \rho \om(\vs) - \frac{1}{2} \left\{ \rho,\om^2 (\vs) \right\}  \right], \quad \om (\vs) = \sum_n \frac{m_n}{(2\pi)^{3/2} r^3_\CSL} \exp \left[ - \frac{(\vs-\ovr_n)^2}{2r^2_\CSL} \right]. 
%\end{align}
\begin{align}
 \cL_\CSL \rho = \frac{\lambda_\CSL}{\pi^{3/2} r^3_\CSL {\rm amu}^2} \int \diff^3 s \left[ \om(\vs) \rho \om(\vs) - \frac{1}{2} \left\{ \rho,\om^2 (\vs) \right\}  \right], \quad \om (\vs) = \sum_n m_n \exp \left[ - \frac{(\vs-\ovr_n)^2}{2r^2_\CSL} \right]. \label{eq:LCSL}
\end{align}
The operator $\om(\vs)$ describes a Gaussian-averaged mass density of the $N$-particle system. In the case of a rigid compound system, the position operator of each particle, $\ovr_n = \ovR + \vr_n^{(0)} + \Delta \ovr_n$, can be expressed in terms of the center-of-mass position operator $\ovR$ of the whole object and $N-1$ relative coordinates. The latter describe the confined motion of the rigidly bound constituents around their equilibrium configuration $\vr_n^{(0)}$ in the center-of-mass system. This motion can be safely neglected, because it is bound to scales well below the CSL localization length $r_\CSL = 100\,$nm (see also \cite{Bassi2003}, Sect.~8.2). We may then write
\begin{align}
\om (\vs) \approx \sum_n m_n \exp \left[ - \frac{\left(\vs-\ovR-\vr_n^{(0)}\right)^2}{2r^2_\CSL} \right] = \frac{r^3_\CSL}{(2\pi)^{3/2}} \int \diff^3 k \, \exp\left[-\frac{r^2_\CSL k^2}{2} \right] e^{i\vk\cdot (\vs-\ovR)} \underbrace{\sum_n m_n e^{-i\vk\cdot \vr_n^{(0)}}}_{\equiv \tilde{\varrho} (\vk)},
\end{align}
introducing the Fourier transform $\tilde{\varrho}(\vk)$ of the object's mass density $\varrho(\vr) = \sum_n m_n \delta \left(\vr - \vr_n^{(0)} \right)$. For the CSL model, the latter can be replaced by the homogeneous mass density of the object, as explained in the main text. 
The CSL master equation (\ref{eq:LCSL}) now acts on the center-of-mass state of motion,
\begin{align}
 \cL_\CSL \rho = \frac{r^3_\CSL \lambda_\CSL}{\pi^{3/2} {\rm amu}^2} \int \diff^3 k\, e^{-r^2_\CSL k^2} \left| \tilde{\varrho} (\vk) \right|^2 \left( e^{i\vk\cdot\ovR} \rho e^{-i\vk\cdot\ovR} - \rho \right).
\end{align}
The exponential operators can be expanded to lowest order in the one-dimensional center-of-mass coordinate $\ox$ in the present case, where the center-of-mass motion is restricted to linear oscillations over amplitudes along the $x$-axis much smaller than $r_\CSL$. This results in the diffusive form $\cL_\CSL \rho \approx -D_\CSL \left[\ox,\left[\ox,\rho\right]\right]/\hbar^2$, with the diffusion rate $D_\CSL = \lambda_\CSL (\hbar/r_\CSL )^2 \alpha$ used in the main text, see Equation (1).

The DP result (11) is obtained analogously after rewriting the DP master equation \cite{Diosi1987} for an object of mass density $\varrho(\vr)$ by means of a Fourier transform,
\begin{align}
 \cL_\DP \rho = -\frac{G}{2\hbar} \int \frac{\diff^3 s_1 \diff^3 s_2}{|\vs_1 - \vs_2|} \left[ \varrho \left( \vs_1 - \ovr \right), \left[ \varrho \left( \vs_2 - \ovr \right), \rho \right] \right] = \frac{G}{2\pi^2\hbar} \int \frac{\diff^3 k}{k^2} \left| \tilde{\varrho} (\vk) \right|^2 \left( e^{i\vk\cdot\ovr} \rho e^{-i\vk\cdot\ovr} - \rho \right).
\end{align}
Note that the Fourier transform of the Coulomb-like term is taken to be the usual $4\pi/k^2$. The diffusion rate $D_\DP$ describes the average growth rate in the second moment of the momentum induced by the above generator $\cL_\DP$. If we are only interested in the one-dimensional motion along the $x$-axis, then $D_\DP = \tr \left( \op_x^2 \cL_\DP \rho \right)$ leads to the expression (11) in the main text.

\section{CSL diffusion for cuboids, spheres and discs}

Here we present the exact expressions for the geometry factors, Eq.~(1) in the main text, of homogeneous rigid bodies of mass $m$ and mean density $\varrho=m/V$. For cuboids of volume $V_{\rm cuboid}=b_x b_y b_z$, discs of volume $V_{\rm disc} = \pi R^2 d$, and spheres of volume $V_{\rm sphere} = 4\pi R^3/3$, the difference lies in the mass density function and its Fourier transform,
\begin{align}
 \tilde{\varrho}_{\rm cuboid} (k_x,k_y,k_z) &= m\,\sinc \left( \frac{k_x b_x}{2} \right)\sinc \left( \frac{k_y b_y}{2} \right)\sinc \left( \frac{k_z b_z}{2} \right), \label{eq:rhoFTcuboid}\\
 \tilde{\varrho}_{\rm disc} (k_x,\vk_\perp) &= \frac{2m}{k_\perp R} J_1 (k_\perp R) \sinc \left( \frac{k_x d}{2} \right), \label{eq:rhoFTdisc}\\
 \tilde{\varrho}_{\rm sphere} (\vk) &= 3m \frac{\sin kR - kR \cos kR}{(kR)^3}. \label{eq:rhoFTsphere}
\end{align}
Here, $J_1$ denotes a Bessel function. After plugging these expressions into the geometry factor (1) in the main text, which determines the momentum diffusion rate for the one-dimensional motion along the $x$-axis, a tedious but straightforward calculation yields
\begin{align}
 \alpha_{\rm cuboid} &= \left( \frac{m}{\rm amu} \right)^2 \Gamma_1 \left( \frac{b_y}{\sqrt{2}r_\CSL} \right) \Gamma_1 \left( \frac{b_z}{\sqrt{2}r_\CSL} \right) \left[1-e^{-b_x^2/4r^2_\CSL} \right]\frac{2r^2_\CSL}{b_x^2}, \label{eq:acuboid}\\
 \alpha_{\rm disc} &= \left( \frac{m}{\rm amu} \right)^2 \Gamma_\perp \left( \frac{R}{\sqrt{2}r_\CSL} \right) \left[1-e^{-d^2/4r^2_\CSL} \right]\frac{2r^2_\CSL}{d^2}, \label{eq:adisc}\\
 \alpha_{\rm sphere} &= \left( \frac{m}{\rm amu} \right)^2 \left[ e^{-R^2/r^2_\CSL} - 1 + \frac{R^2}{2r^2_\CSL} \left( e^{-R^2/r^2_\CSL} + 1 \right) \right] \frac{6r^6_\CSL}{R^6}, \label{eq:asphere}
\end{align}
with the abbreviations
\begin{align}
 \Gamma_1 (\xi) &= \frac{2}{\xi^2} \left[ e^{-\xi^2/2} - 1 + \sqrt{\frac{\pi}{2}}\xi \erf \left( \frac{\xi}{\sqrt{2}} \right) \right], & \Gamma_\perp (\xi) = \frac{2}{\xi^2} \left\{ 1 - e^{-\xi^2} \left[ I_0 (\xi^2) + I_1 (\xi^2) \right] \right\}.
\end{align}
The terms $I_{0,1}$ denote modified Bessel functions. Equations (2) and (3) in the main text are obtained by expanding the exact geometry factors in $R/r_\CSL$, $b/r_\CSL$, and $d/r_\CSL$ asymptotically.

\section{DP diffusion for cubic crystal lattices}

Here we calculate the DP diffusion rate, Eq.~(11) in the main text, for macroscopic solids consisting of a cubic and mono\-atomic crystal lattice. For simplicity, we neglect the small electron mass and assume that the nuclear mass $m_{\rm A}$ at each lattice point is on average distributed evenly according to the Gaussian mass density distribution $\varrho_{\rm A} (\vr) = m_{\rm A} \exp (-r^2/2\sigma_\DP^2)/(2\pi\sigma_\DP^2)^{3/2}$, with $\tilde{\varrho}_{\rm A} (\vk) = m_{\rm A}\exp (-\sigma_\DP^2 k^2 /2)$ its Fourier transform. The spread $\sigma_\DP$ is assumed to be much smaller than the lattice constant $a$, so that the average mass densities of neighboring lattice points do not overlap.

The total mass density of the object and its Fourier transform can now be written as
\begin{align}
 \varrho (\vr) &= \chi \left( \vr \right) \sum_{j,n,\ell= -\infty}^{\infty}\varrho_{\rm A} (x - ja, y - na, z - \ell a), \nonumber \\
 \tilde{\varrho} (\vk) &= \frac{1}{a^3} \sum_{j,n,\ell} \tilde{\varrho}_{\rm A} \left( \frac{2\pi j}{a},\frac{2\pi n}{a},\frac{2\pi \ell}{a} \right) \tilde{\chi} \left( k_x - \frac{2\pi j}{a}, k_y - \frac{2\pi n}{a}, k_z - \frac{2\pi \ell}{a} \right)
\end{align}
where $\chi(\vr)$ denotes the characteristic function of the given body shape; it is unity for all points inside the body volume and zero elsewhere (i.e.~$\chi$ is proportional to the homogeneous mass density employed in the CSL case). Its Fourier transform is denoted by $\tilde{\chi}$. The function $\chi (\vr)$ varies on essentially macroscopic scales, whereas the lattice sum is a sharply peaked periodic function oscillating on the microscopic scales $\sigma_\DP,a$. Given the macroscopic volume of the object, $V^{1/3} \gg a \gg \sigma_\DP$, the Fourier transform $\tilde{\chi}$ of the characteristic function has a width of the order of $V^{-1/3}$, much smaller than $2\pi/a$. Hence, we may reduce the double summation to a single sum when taking the absolute square of $\tilde{\varrho}(\vk)$ and write
\begin{align}
 \left| \tilde{\varrho} (\vk) \right|^2 \approx \frac{1}{a^6} \sum_{j,n,\ell} \left| \tilde{\varrho}_{\rm A} \left( \vG_{jn\ell}\right) \tilde{\chi} \left( \vk - \vG_{jn\ell} \right) \right|^2,
\end{align}
with $\vG_{jn\ell} = 2\pi \left( j,n,\ell \right)/a$ a reciprocal lattice vector.
Plugging this into the DP diffusion rate (11) and exploiting once again the sharply peaked nature of $\tilde{\chi}$, we arrive at
\begin{align}
 D_\DP &= \frac{G \hbar}{2\pi^2 a^6} \sum_{jn\ell} \left| \tilde{\varrho}_{\rm A} \left( \vG_{jn\ell}\right) \right|^2 \int \diff^3 k \, \frac{k_x^2}{\vk^2} \left| \tilde{\chi} \left( \vk - \vG_{jn\ell} \right) \right|^2 \approx \frac{G \hbar}{2\pi^2 a^6} \sum_{jn\ell} \left| \tilde{\varrho}_{\rm A} \left( \vG_{jn\ell}\right) \right|^2 \frac{j^2}{j^2+n^2+\ell^2} \int\diff^3 k \, \left| \tilde{\chi} \left( \vk \right) \right|^2.
\end{align}
The latter integral can be evaluated using $\chi^2 = \chi$, that is, $\int\diff^3 k\, |\tilde{\chi}(\vk)|^2 = (2\pi)^3 \int\diff^3 r \, \chi^2 (\vr) = (2\pi)^3 V$. Moreover, the function $\tilde{\varrho}_{\rm A} (\vG_{jn\ell})$ extends over many reciprocal lattice vectors, since $1/\sigma_\DP \gg 2\pi/a$. This allows us to approximate the lattice sum by an integral,
\begin{align}
 D_\DP &\approx \frac{4\pi G \hbar V}{a^6} \left( \frac{a}{2\pi} \right)^3 \int\diff^3 q \, \frac{q_x^2}{\vq^2} \left| \tilde{\varrho}_{\rm A} (\vq) \right|^2 = \frac{G \hbar m_{\rm A}^2 V}{6\pi^2 a^3} \int \diff^3 q \, e^{-\sigma_\DP^2 q^2} = \frac{G \hbar m_{\rm A}^2 V}{6\sqrt{\pi} a^3 \sigma_\DP^3}
\end{align}
Noting that $m = m_{\rm A} V/a^3$ is the total mass of the object and $\varrho = m_{\rm A}/a^3$ its mean density, we arrive at the result (12) given in the main text.

\end{document}